\documentclass[12pt]{article}

\usepackage{tocloft}
\usepackage{graphicx}
\usepackage{natbib}
\usepackage[protrusion=true,expansion=true]{microtype}
\usepackage{amsmath,amsthm,amsfonts,amssymb,mathrsfs,dsfont,mathtools}
\usepackage{booktabs}
\usepackage{url}
\usepackage{float}
\usepackage{color}
\usepackage{parskip}
\usepackage[explicit]{titlesec}
\usepackage{threeparttable}
\usepackage{setspace}
\usepackage[margin=0.85in]{geometry}
\usepackage[font = onehalfspacing]{caption}
\usepackage[hyperfootnotes=true]{hyperref}
\usepackage[labelfont=bf]{caption}
\usepackage{authblk}
\usepackage{caption}
\usepackage[titletoc,toc,title]{appendix}
\usepackage{enumerate}
\usepackage{soul}
\usepackage{subfig}
\usepackage{accents}
\usepackage{subfiles} 
\usepackage[capitalise, nameinlink]{cleveref}
\usepackage{dsfont}

\usepackage{changepage}
\usepackage[noend]{algpseudocode}
\usepackage{tocloft}
\usepackage[protrusion=true,expansion=true]{microtype}
\usepackage{amsmath,amsthm,amsfonts,amssymb,mathrsfs,dsfont,mathtools}
\usepackage{float}
\usepackage{color}
\usepackage{parskip}
\usepackage[explicit]{titlesec}
\usepackage{threeparttable}
\usepackage{setspace}
\usepackage[font = onehalfspacing]{caption}
\usepackage[usenames,dvipsnames]{xcolor}
\usepackage[hyperfootnotes=true]{hyperref}
\usepackage[labelfont=bf]{caption}
\usepackage{authblk}
\usepackage[titletoc,toc,title]{appendix}
\usepackage{enumerate}
\usepackage[capitalise, nameinlink]{cleveref}
\usepackage{diagbox}
\usepackage{array,multirow}

\usepackage[utf8]{inputenc} 
\usepackage[T1]{fontenc}    
\usepackage{hyperref}       
\usepackage{url}            
\usepackage{booktabs}       
\usepackage{amsfonts}       
\usepackage{nicefrac}       
\usepackage{microtype}      
\usepackage{lipsum}		
\usepackage{graphicx}
\usepackage{natbib}
\usepackage{doi}
\usepackage{subfiles}
\usepackage{hyphenat}

\usepackage{adjustbox}
\usepackage{multirow}
\usepackage{amsmath}
\usepackage[section]{placeins}
\usepackage{algorithm}
\usepackage{algpseudocode}
\usepackage{dsfont}

\DeclareUnicodeCharacter{2212}{-}

\setstcolor{red} 

\numberwithin{equation}{section}

\hypersetup{colorlinks = true, citecolor = MidnightBlue, linkcolor = MidnightBlue, urlcolor = MidnightBlue}

\titleformat{\section}{\normalfont\large\bfseries}{\thesection}{1em}{#1}
\titleformat{\subsection}{\normalfont\normalsize\bfseries}{\thesubsection}{1em}{#1}
\titleformat{\subsubsection}{\normalfont\normalsize\itshape}{\thesubsubsection}{1em}{#1}
\titlespacing\section{0pt}{12pt plus 4pt minus 2pt}{6pt plus 2pt minus 2pt}
\titlespacing\subsection{0pt}{12pt plus 4pt minus 2pt}{3pt plus 2pt minus 3pt}
\titlespacing\subsubsection{0pt}{12pt plus 4pt minus 2pt}{0pt plus 2pt minus 3pt}

\def\boxit#1{\vbox{\hrule\hbox{\vrule\kern6pt
          \vbox{\kern6pt#1\kern6pt}\kern6pt\vrule}\hrule}}

\definecolor{orange}{rgb}{1,0.5,0}
\definecolor{MyDarkBlue}{rgb}{0,0.08,0.45}

\begin{document}
\title{Is Deep Hedging Reinforcement Learning?\\
\normalsize A Response to the View that Monte Carlo-Based Policy Gradient Methods\\ Fall Outside the Reinforcement Learning Framework\thanks{I thank the Natural Sciences and Engineering Research Council of Canada (Godin: RGPIN-2024-04593) for their financial support.}}
\author[,a,b]{Fr\'ed\'eric Godin\thanks{Corresponding author. \vspace{0.2em} \newline
{\mbox{\hspace{0.47cm}} \it Email address:} 
\href{mailto:frederic.godin@concordia.ca}{frederic.godin@concordia.ca} (Fr\'ed\'eric Godin).  }}
\affil[a]{{\small Concordia University, Department of Mathematics and Statistics, Montr\'eal, Canada}}
\affil[b]{{\small Quantact Laboratory, Centre de Recherches Math\'ematiques, Montr\'eal, Canada}}

\vspace{-25pt}
\date{ 
\today}


\maketitle \thispagestyle{empty} 

%

\vspace{-15pt}

\begin{abstract}
\vspace{-5pt}

{\footnotesize    
The deep hedging framework of \citet{buehler2019deep} trains a neural network policy, via Monte Carlo simulation of price paths and stochastic gradient descent, to minimize a risk measure applied to the terminal hedging error. In a recent stream of papers, my coauthors and I have described this technique as reinforcement learning (RL). Several peers have, on occasion, expressed the view that deep hedging does not constitute genuine RL, on two grounds, among others: first, that because feedback is generated only at the terminal date, with no intermediate reward signal, the method cannot constitute genuine RL; and second, that the absence of a value function, a Bellman equation, temporal-difference (TD) learning, and an explicit exploration mechanism disqualifies the method from the RL category altogether, so that it should instead be labeled a neural-network method for stochastic optimal control. The present note argues instead that both objections rest on an unduly narrow, TD-centric reading of what constitutes RL, and that once RL is understood, as it is in the standard references of the field, to include Monte Carlo policy-gradient methods and direct (actor-only) policy search as first-class members, the deep hedging algorithm of \citet{buehler2019deep} falls squarely within the RL umbrella. 
} 
	
\bigskip 



\noindent \textbf{Keywords:} Deep Hedging, Reinforcement Learning, Machine Learning Taxonomy.
\end{abstract}

\medskip

\thispagestyle{empty} \vfill \pagebreak

\setcounter{page}{1}
\pagenumbering{roman}

\doublespacing

\setcounter{page}{1}
\pagenumbering{arabic}


\section{Introduction}\label{se:intro}

Deep hedging, as introduced by \citet{buehler2019deep}, trains a neural network policy to map state variables to a hedging portfolio rebalancing decision, by simulating a large number of price paths, computing a risk measure of the resulting terminal hedging error under the candidate policy, and updating the network's parameters by (stochastic) gradient descent, with gradients obtained by backpropagating through the simulated paths. In a recent stream of papers, see for instance \cite{franccois2024enhancing,neagu2024deep,franccois2025difference,franccois2025deep,ahmadi2025learning,neagu2025deep, perez2026hedging}, my coauthors and I have described this approach as an instance of reinforcement learning (RL). An opinion shared among some peers is that this characterization is mistaken, on essentially two grounds.

The first is that the learning signal in deep hedging is generated only once, at the terminal date of each simulated path, and that no reward or feedback is produced at intermediate rebalancing dates; a method that only ever learns from a single end-of-episode outcome, on this view, does not perform the kind of sequential, step-by-step learning that RL is supposed to involve. The second is that deep hedging, as implemented, contains none of the algorithmic machinery usually associated with RL -- no value function, no Bellman equation, no temporal-difference (TD) updates, and no explicit exploration--exploitation mechanism -- and should therefore be described more modestly as a direct, simulation-based numerical scheme for solving a stochastic control problem, rather than as RL.

Both of these arguments correctly identify a real feature of deep hedging: it is a Monte Carlo, actor-only, pathwise-gradient method, and not a temporal-difference or value-based one. Where the present note departs is in the inference drawn from that feature, namely that it places deep hedging outside reinforcement learning. \cref{sec:Rldef} revisits the definition of RL given by \citet{suttonbarto2018}, whose textbook is widely regarded as a standard reference in the field. This foundation is used in \cref{se:response} to directly address the two arguments described above and explain why deep hedging is best understood as falling under the RL umbrella.


\section{Reinforcement Learning According to \citet{suttonbarto2018}}\label{sec:Rldef}

\citet{suttonbarto2018} frame reinforcement learning as a way for an agent to learn, through interacting with an environment, which actions to take in order to maximize a cumulative reward signal -- discovering good behavior through trial and error rather than being told the correct action. They frame the underlying problem as a Markov decision process (MDP) -- an agent interacting with an environment through states, actions, and rewards over time -- and identify exploration versus exploitation as a tension intrinsic to the problem, since the agent must balance acting on what it already believes to be good against trying alternatives that might be better.

Crucially, \citet{suttonbarto2018} organize the algorithms that solve this problem along two largely independent axes, neither of which is part of the definition of RL itself. The first axis is \emph{value-based versus policy-based}: some methods learn an explicit value function and derive a policy from it, others learn a policy directly with no value function at all. The second axis is \emph{Monte Carlo versus temporal-difference}: some methods update only once an entire episode has been simulated to completion, using the single realized return for that episode, while others bootstrap from an intermediate value estimate at every step. \citet{suttonbarto2018} present Monte Carlo control -- simulating a full episode and updating from its terminal return alone -- as a foundational RL method in its own right, prior to introducing TD learning, precisely because full-episode, terminal-return updating is already a valid and complete solution method for an MDP, with no bootstrapping required. Direct policy search, of which the REINFORCE algorithm of \citet{williams1992} is the canonical example, sits at the intersection of these two features: it parametrizes a policy directly, simulates full trajectories, and updates parameters from the trajectory's realized return using the policy-gradient theorem, without ever constructing a value function, a Bellman equation, or a TD update. REINFORCE is, and always has been, considered a bona fide RL algorithm.

This organization reflects a broader point about how \citet{suttonbarto2018} define the field: RL is defined by the problem it studies, not by any specific algorithm used to solve it. The problem is that of an agent learning, through interaction, to achieve a goal; value functions, policies, and models of the environment are each just tools that different classes of RL algorithms -- Monte Carlo methods, TD learning, policy-gradient methods, and others -- may or may not employ in solving that problem. It follows that no single one of these tools, including a value function, a Bellman equation, or TD updates, is part of the definition of RL itself; a method's use or non-use of any particular tool bears on which class of RL algorithm it belongs to, not on whether it belongs to RL at all.


\section{Deep Hedging as a Monte Carlo Policy-Gradient Method}\label{se:response}

Read against this taxonomy, neither of these two arguments against describing deep hedging as RL holds up.
On the first point, feedback confined to the terminal date of a simulated path is not a departure from reinforcement learning; it is precisely the Monte Carlo credit-assignment structure described above, in which the entire episode's return is propagated back to every decision along the trajectory in a single update, rather than incrementally at each step as under TD learning. The propagation is carried out by automatic differentiation through each simulated path, making it a pathwise-derivative estimator — a derivative computed along each realized trajectory and then averaged across a batch of paths,\footnote{When the training objective is the sample mean of the terminal loss, the batch gradient is literally the average of the per-path pathwise derivatives. For nonlinear risk measures such as CVaR or entropic risk, the objective is instead a functional of the empirical distribution of the batch's terminal losses (e.g., an average over the worst order statistics, or a log-sum-exp of the losses), so the gradient is obtained by applying the chain rule through that functional — via a subgradient at the empirical quantile for CVaR, or directly for smooth functionals like entropic risk — rather than by a simple unweighted average of per-path derivatives. The gradient remains a Monte Carlo, pathwise estimator in either case: each per-path terminal loss is still differentiated via backpropagation through its simulated path, and it is only the way per-path derivatives are combined into a batch-level gradient that depends on the choice of risk measure.} rather than a single-path computation — in contrast to the score-function (also called likelihood-ratio) estimator used by REINFORCE, which reweights sampled outcomes by the log-derivative of the sampling distribution instead of differentiating the outcome itself. Both are Monte Carlo estimators of a policy gradient from full-episode returns, differing only in which of the two standard estimators is used. This pathwise-gradient family is studied under the name (stochastic) value gradients methods by \citet{heess2015learning}.
On the second point, the absence of a value function, a Bellman equation, or TD updates is exactly what one would expect of a policy-based, Monte Carlo method, and \cref{sec:Rldef} shows that such methods are a first-class part of the RL taxonomy, not an exception to it. REINFORCE and the critic-free, full-trajectory variants of stochastic value gradients \citep{heess2015learning} share this same absence; other pathwise-gradient methods, such as deterministic policy gradient \citep{silver2014deterministic} and DDPG \citep{lillicrap2016continuous}, instead pair a pathwise-gradient actor update with a learned critic (value function). If anything, this reinforces the point: the same policy-gradient family spans both critic-free and critic-based variants, and neither is considered to fall outside RL on that account.
The absence of an explicit exploration mechanism is similarly non-disqualifying. Explicit exploration is required when the effect of alternative actions on future outcomes is initially unknown and must be discovered through interaction with the environment. This need does not arise in deep hedging. Because the price dynamics are either known and differentiable or, when training on historical data, treated as exogenous to the hedging decisions, the policy gradient can be estimated directly by differentiating through sampled trajectories, without requiring trial-and-error exploration of alternative actions. Model-based policy-gradient methods trained against known or learned dynamics models, such as those of \citet{heess2015learning}, may dispense with explicit exploration noise for the same reason.
It is also worth noting that the RL literature applied to hedging and related control problems already spans both ends of the taxonomy described in \cref{sec:Rldef}: value-based TD formulations, such as the Q-learning approach of \citet{kolmritter2019dynamic}, coexist with policy-gradient, pathwise-gradient formulations closer to deep hedging. Treating the presence of a value function or TD update as a necessary condition for RL would place these two published, RL-labeled approaches to essentially the same control problem on opposite sides of the boundary, which is not a distinction the field itself draws. It is more consistent with standard usage to treat both as RL methods occupying different corners of the same taxonomy — TD/value-based versus Monte Carlo/policy-gradient — than to treat only one of them as genuine RL.


\section{Concluding Remarks}

The view that \cite{buehler2019deep}'s deep hedging approach is a Monte Carlo, policy-gradient method with no value function, Bellman equation, TD update, or explicit exploration mechanism is correct, and is not in dispute here. What is at issue is whether this observation implies that deep hedging is not RL. Once reinforcement learning is defined, as it is by \citet{suttonbarto2018}, as the general problem of learning a policy from interaction to maximize cumulative reward (or minimize its associated risk), and once its algorithms are recognized as spanning both value-based and policy-based methods, and both temporal-difference and full-episode Monte Carlo updating, deep hedging falls squarely within it.
Deep hedging is therefore most accurately described as a Monte Carlo, actor-only, pathwise-gradient policy-gradient reinforcement learning algorithm. This terminology is both technically precise and fully consistent with the standard reinforcement learning literature.

\section*{Acknowledgements}

I gratefully acknowledge the assistance of Claude AI (Anthropic) and ChatGPT (OpenAI) in the writing of this note -- a modern way, in my view, of standing on the shoulders of giants. I have no conflict of interest to declare.


\bibliographystyle{apalike}
\bibliography{references}

@book{suttonbarto2018,
  author    = {Sutton, Richard S. and Barto, Andrew G.},
  title     = {Reinforcement Learning: An Introduction},
  edition   = {2nd},
  publisher = {MIT Press},
  year      = {2018}
}

@article{williams1992,
  author  = {Williams, Ronald J.},
  title   = {Simple Statistical Gradient-Following Algorithms for Connectionist Reinforcement Learning},
  journal = {Machine Learning},
  volume  = {8},
  number  = {3--4},
  pages   = {229--256},
  year    = {1992}
}

@article{buehler2019deep,
  author  = {Buehler, Hans and Gonon, Lukas and Teichmann, Josef and Wood, Ben},
  title   = {Deep Hedging},
  journal = {Quantitative Finance},
  volume  = {19},
  number  = {8},
  pages   = {1271--1291},
  year    = {2019}
}

@article{kolmritter2019dynamic,
  author  = {Kolm, Petter N. and Ritter, Gordon},
  title   = {Dynamic Replication and Hedging: A Reinforcement Learning Approach},
  journal = {The Journal of Financial Data Science},
  volume  = {1},
  number  = {1},
  pages   = {159--171},
  year    = {2019}
}

@inproceedings{heess2015learning,
  author    = {Heess, Nicolas and Wayne, Greg and Silver, David and Lillicrap, Timothy and Erez, Tom and Tassa, Yuval},
  title     = {Learning Continuous Control Policies by Stochastic Value Gradients},
  booktitle = {Advances in Neural Information Processing Systems (NeurIPS)},
  volume    = {28},
  year      = {2015}
}

@inproceedings{silver2014deterministic,
  author    = {Silver, David and Lever, Guy and Heess, Nicolas and Degris, Thomas and Wierstra, Daan and Riedmiller, Martin},
  title     = {Deterministic Policy Gradient Algorithms},
  booktitle = {Proceedings of the 31st International Conference on Machine Learning (ICML)},
  year      = {2014}
}

@inproceedings{lillicrap2016continuous,
  author    = {Lillicrap, Timothy P. and Hunt, Jonathan J. and Pritzel, Alexander and Heess, Nicolas and Erez, Tom and Tassa, Yuval and Silver, David and Wierstra, Daan},
  title     = {Continuous Control with Deep Reinforcement Learning},
  booktitle = {International Conference on Learning Representations (ICLR)},
  year      = {2016}
}

@article{franccois2025deep,
  title={Deep hedging with options using the implied volatility surface},
  author={Fran{\c{c}}ois, Pascal and Gauthier, Genevi{\`e}ve and Godin, Fr{\'e}d{\'e}ric and P{\'e}rez-Mendoza, Carlos O},
  journal={arXiv preprint arXiv:2504.06208},
  year={2025}
}

@article{franccois2025difference,
  title={Is the difference between deep hedging and delta hedging a statistical arbitrage?},
  author={Fran{\c{c}}ois, Pascal and Gauthier, Genevi{\`e}ve and Godin, Fr{\'e}d{\'e}ric and Mendoza, Carlos Octavio P{\'e}rez},
  journal={Finance Research Letters},
  volume={73},
  pages={106590},
  year={2025},
  publisher={Elsevier}
}

@article{franccois2024enhancing,
  title={Enhancing deep hedging of options with implied volatility surface feedback information},
  author={Fran{\c{c}}ois, Pascal and Gauthier, Genevi{\`e}ve and Godin, Fr{\'e}d{\'e}ric and Mendoza, Carlos Octavio P{\'e}rez},
  journal={arXiv preprint arXiv:2407.21138},
  year={2024}
}

@article{ahmadi2025learning,
  title={Learning to Hedge Swaptions},
  author={Ahmadi, Zaniar and Godin, Fr{\'e}d{\'e}ric},
  journal={arXiv preprint arXiv:2512.06639},
  year={2025}
}

@article{perez2026hedging,
  title={Hedging targeted risks with reinforcement learning: application to life insurance contracts with embedded guarantees},
  author={P{\'e}rez-Mendoza, Carlos Octavio and Godin, Fr{\'e}d{\'e}ric},
  journal={ASTIN Bulletin: The Journal of the IAA},
  volume={56},
  number={2},
  pages={420--446},
  year={2026},
  publisher={Cambridge University Press}
}

@inproceedings{neagu2024deep,
  title={Deep Hedging with Market Impact.},
  author={Neagu, Andrei and Godin, Fr{\'e}d{\'e}ric and Simard, Clarence and Kosseim, Leila and others},
  booktitle={Canadian AI},
  year={2024}
}

@article{neagu2025deep,
  title={Deep Reinforcement Learning Algorithms for Option Hedging},
  author={Neagu, Andrei and Godin, Fr{\'e}d{\'e}ric and Kosseim, Leila},
  journal={arXiv preprint arXiv:2504.05521},
  year={2025}
}

\end{document}